\documentclass[twoside,onecolumn]{amsart}
\usepackage{DejaVuSans}
\usepackage[T1]{fontenc}

\usepackage[english]{babel} 

\usepackage[hmarginratio=1:1,top=28mm,bottom=28mm,left=21mm,right=21mm,footskip=10mm]{geometry} 
\usepackage[hang, small,labelfont=bf,up,textfont=it,up]{caption} 
\usepackage{booktabs} 

\usepackage{abstract} 

\usepackage{enumerate}

\usepackage[]{titlesec}
\titleformat{\section}[block]
  {\fontsize{11}{10}\bfseries\sffamily}
  {\thesection}
  {1em}
  {}
\titleformat{\subsection}[block]
  {\fontsize{11}{10}\bfseries\sffamily}
  {\thesubsection} 
  {1em}
  {}
  
\usepackage{fancyhdr} 
\usepackage{titling} 
\usepackage[noblocks]{authblk}

\usepackage{hyperref} 
\usepackage{natbib}

\usepackage{url,lineno,microtype,subcaption}
\usepackage[onehalfspacing]{setspace}

\usepackage{inputenc,crop,graphicx,amsmath,array,color}
\usepackage{amssymb,flushend,stfloats,amsthm,chngpage,parskip,epstopdf}
\usepackage{float}
\usepackage[htt]{hyphenat}

\pretitle{\begin{center}\vspace{10pt}\LARGE\bfseries} 
\posttitle{\vspace{16pt}\end{center}} 

\title{Black Hole Mass Estimation in Type 1 AGN: H$\beta$ vs. Mg II lines and the role of Balmer continuum} 
\date{}
\author[1]{Jelena Kova{\v c}evi\' c-Doj{\v c}inovi\' c}
\author[1,2,*]{Sladjana Mar{\v c}eta-Mandi\' c}
\author[1,2,*]{Luka {\v C}.~Popovi\' c}

\begin{tiny}
\affil[1]{Astronomical Observatory, Volgina 7, Belgrade, Serbia}
\affil[2]{Department of Astronomy, Faculty of Mathematics, University of Belgrade, Studentski trg 16, Belgrade, Serbia}
\affil[*]{sladjana@aob.rs   //   lpopovic@aob.rs} 

\end{tiny}

\begin{document}
\maketitle\date{\vspace{-10ex}}

\vspace{-2ex}
\begin{abstract}
Here we investigate the H$\beta$ and Mg II spectral line parameters used for the black hole mass (M$_{\rm BH}$) estimation for a sample of Type 1 Active Galactic Nuclei (AGN) spectra selected from the Sloan Digital Sky Survey (SDSS) database. We have analyzed and compared the virialization of the H$\beta$ and Mg II emission lines, and found that the H$\beta$ line is more confident virial estimator than Mg II. We have investigated the influence of the Balmer continuum emission to the M$_{\rm BH}$ estimation from the UV parameters, and found that the Balmer continuum emission can contribute to the overestimation of the M$_{\rm BH}$ on average for $\sim$ 5$\%$ (up to 10$\%$).

\vspace{6pt}\small{\bf keywords}: galaxies: active, galaxies: nuclei, quasars: supermassive black holes, techniques: spectroscopic, quasars: emission lines, line: profiles
\end{abstract}



\section{Introduction}
Several methods are used to estimate central black hole (BH) mass M$_{\rm BH}$ in galaxies \citep[for review see e.g.][]{Marziani_Sulentic_2012,Shen_2013,Peterson_2014,Ilic_Popovic_2014}. For Type 1 AGN, the most appropriate methods for the M$_{\rm BH}$ estimation are those using the strong broad emission lines (BELs), as the most prominent features in their spectra. The virial methods \citep[see][] {Peterson_2004,Vestergaard_Peterson_2006} are based on the assumption that the Broad Line Region (BLR) gas is bounded to the central BH \citep[see][]{Gaskell_2009} and the main broadening mechanism of the BELs is the Keplerian motion around the supermassive BH, so the full width at half maximum (FWHM) of BELs indicates the velocity of the emitting gas. We should note that in principle the line dispersion much better represents this motion \citep[][]{Peterson_2004,Collin_2006}, however in order to find the line dispersion one should assume some type of line profile (that may be very complex), therefore the FWHM is often used instead of the line dispersion.

One of these methods is based on the R - L relationship \citep[see e.g.][]{Bentz_2006}, the outcome of the reverberation mapping \citep[see][etc.]{Blandford_McKee_1982,Peterson_2004}, which enables the estimation of the photometric radius from only one epoch spectrum \citep[see e.g.][]{Vestergaard_Peterson_2006}. An alternative method for M$_{\rm BH}$ estimation using the BEL parameters is based on the gravitational redshift in the broad line profiles \citep[see][]{Zheng_Sulentic_1990,Popovic_1995, Bon_2015,Jonic_2016,Liu_2017}. The advantage of this method is that it does not depend on the BLR inclination, unlike the virial methods.

There are many unresolved questions relevant for the application of these methods. For example, since the BLR geometry could be complex \citep[see e.g.][etc.]{Sulentic_2000,Popovic_2004,Gaskell_2009}, it is essential to confirm if the virial assumption is correct for all BELs which are used in the methods for the  M$_{\rm BH}$ estimation and if the gravitational redshift could be measured from the BELs complex shapes, or if it may be suppressed by some other effects.

The most frequently used BELs as the virial estimators are the broad H$\beta$ (in the optical) and Mg II (in the UV) lines \citep[see][]{Marziani_Sulentic_2012}. Both, H$\beta$ and Mg II lines have complex profiles, which should be considered if these lines are used for the M$_{\rm BH}$ estimation. Extracting refined H$\beta$ and Mg II profiles is a difficult task and it is essential for an accurate M$_{\rm BH}$ estimation. Especially since the broad H$\beta$ overlaps with a numerous optical Fe II lines, the [O III] doublet and the H$\beta$ narrow line component, while the Mg II line overlaps with a numerous UV Fe II lines. Finally, the presence of the Balmer continuum for $\lambda<$ 3646\AA, is contributing to the uncertainty of the M$_{\rm BH}$ estimation from the UV parameters and it has to be subtracted for obtaining the pure power law luminosity in the UV band.

In this paper we first present the models of the optical Fe II, UV Fe II emission and Balmer continuum, that could give more precise measurements of the optical and UV parameters (H$\beta$ and Mg II broad line profiles, power law luminosity at $\lambda$ = 3000\AA, as L$_{\lambda}(3000\mathring A)$) used for the M$_{\rm BH}$ estimation. Then, we analyze the virialization assumption for the H$\beta$ and Mg II broad lines, and the influence of the Balmer continuum to the M$_{\rm BH}$ estimation from the UV parameters.


\section{The Sample and Analysis}

The used sample consists of the 287 spectra of Type 1 AGN, obtained from the SDSS Data Release 7 (DR7). The sample is the same as in \citet{Kovecevic-Dojcinovic_2015} where the detailed description of the sample selection criteria is given. For the investigation of the virialization of the emission regions, we exclude all spectra with the blue asymmetry, which resulted with the sample of 123 objects used in this research of the H$\beta$ and Mg II profiles \citep[see][]{Jonic_2016}. In the future work we plan to investigate in more details radio properties for this sample, and to search for the connection between the radio-loudness and M$_{\rm BH}$.

\subsection{Model of the Optical Emission Lines in 4000-5500{\AA}: Extracting the Pure Broad H$\beta$ Profile}

To obtain a pure broad H$\beta$ component, the narrow H$\beta$ and [O III] lines have to be carefully subtracted, as well as the optical Fe II lines.   
After correcting the spectra for the Galactic reddening and the cosmological redshift, and subtracting the underlying continuum, we applied the multi-Gaussian fitting procedure in 4000-5500{\AA} range, described in details in \citet{Kovacevic_2010} and \citet{Kovecevic-Dojcinovic_2015}. In the fitting procedure, the number of free parameters was reduced assuming that the lines or the line components which originate from the same emission region, have the same widths and shifts. Therefore all narrow Balmer lines (H$\delta$, H$\gamma$ and H$\beta$) have the same widths and shifts as [O III] lines, since we assume that they all originate from the Narrow Line Region (NLR). The broad part of the Balmer lines was modeled with two Gaussian functions representing the emission from the Intermediate Line Region (ILR) and from the Very Broad Line Region (VBLR) \citep[see][]{Popovic_2004,Bon_2006,Bon_2009_MNRAS,Hu_2008}. 

\begin{figure*}[h]
\begin{center}
\centering\includegraphics[width=17.5cm]{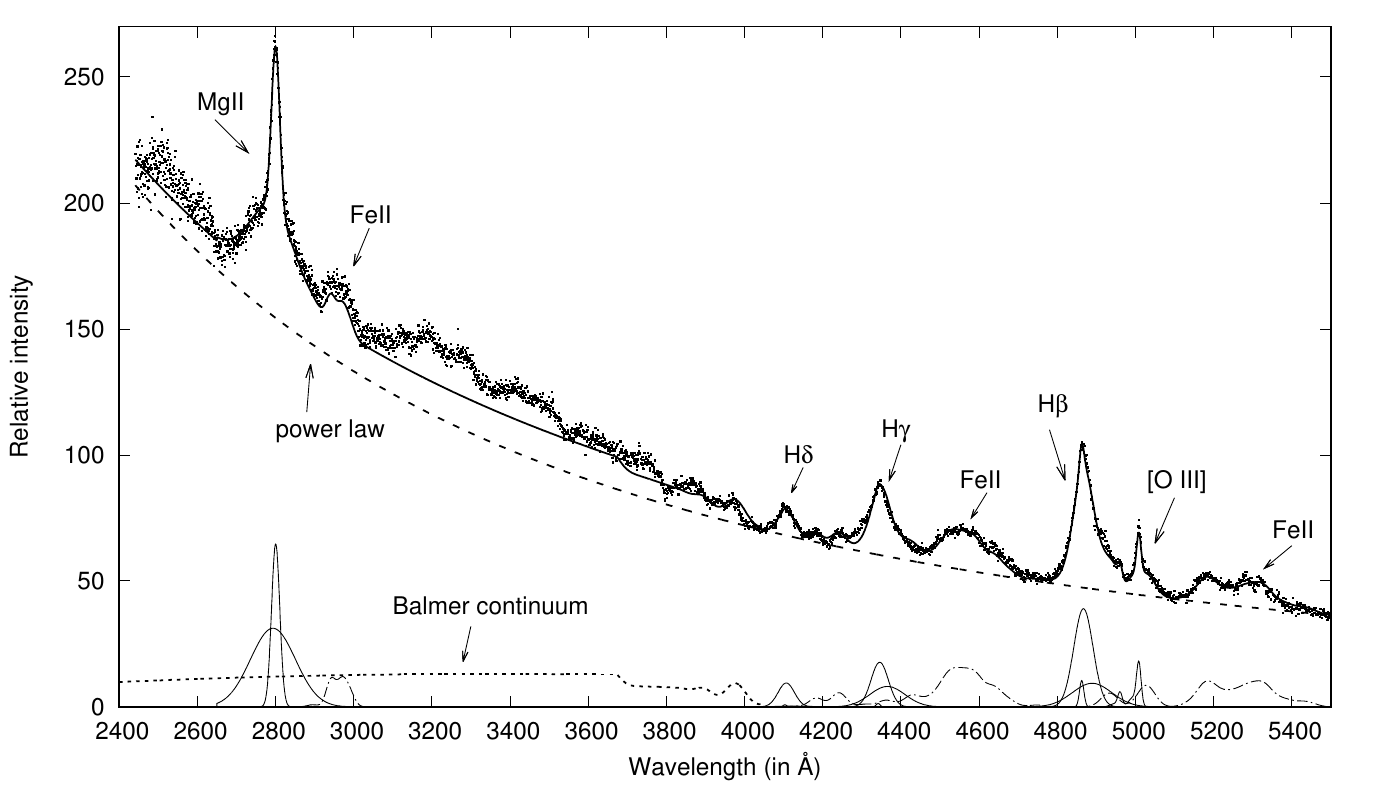}
\caption{The example of the decomposition of the UV-optical pseudo-continuum (power law + Balmer continuum) and emission lines in ranges 2650-3050{\AA} and 4000-5500{\AA} for object SDSS J014942.50+001501.7. The observations are denoted with dots, the sum of pseudo-continuum and emission line model with solid line, and the power law continuum with dashed line. Bellow, the Balmer continuum is given with dotted line, and UV/optical Fe II templates with dotted-dashed line and all emission lines with solid line. }
\label{fig:1}
\end{center}
\end{figure*}

We have made a Fe II template as a sum of the most prominent Fe II lines, described with Gaussian functions, with the same widths and shifts, since we assumed that all optical Fe II lines were originating from the same emission region. The number of free parameters was reduced by calculating the relative intensities for the Fe II lines with the same lower term of transition \citep[see][]{Kovacevic_2010}. Finally, the Fe II template was described with five parameters of intensity, width, shift and temperature, which was included in the calculation of the relative intensities. For more details about the optical Fe II template see \citet{Kovacevic_2010} and \citet{Shapovalova_2012}, and  this Fe II template is also available on line $(http://servo.aob.rs/FeII{\_}AGN)$. The example of the spectral decomposition in 4000-5500{\AA} range is shown in the optical part in Figure 1.

\subsection{The Balmer Continuum Model}
The UV pseudo-continuum consists of the power law, which represents the emission from the accretion disc and the bump at 3000\AA, which represents the sum of the blended, high-order broad Balmer lines and the Balmer continuum ($\lambda<$ 3646{\AA}). In order to measure the flux or luminosity of the power law at UV spectral range (e.g. 3000\AA), one needs to subtract the Balmer continuum emission first (see Figure 1).
The model of the Balmer continuum first given in \citet{Kovacevic_2014}, is based on the function for the Balmer continuum given in \citet{Grandi_1982} for the case of a partially optically thick cloud, with one degree of freedom decreased, as the intensity of the Balmer continuum was calculated, obtaining in that way lower uncertainty. 
The intensity of the Balmer continuum was estimated at the Balmer edge ($\lambda$ = 3646\AA), as a sum of the intensities of all high-order Balmer lines at the same wavelength. All broad Balmer lines were represented with one Gaussian function only, with the same width and shift of a prominent Balmer line, and their relative intensities were taken from the literature or were calculated \citep[see][]{Kovacevic_2014}. Therefore, if only one prominent Balmer line was fitted (e.g. H$\beta$), and the shift, width and intensity were obtained from that fit, than the fluxes of all other Balmer lines would be known, and the Balmer continuum at the Balmer edge could be calculated. Finally, with this model, the UV pseudo-continuum was fitted with four free parameters: the width, shift and intensity of the one prominent Balmer line, and the exponent of the power law. An example of the Balmer continuum fit is shown below in Figure 1.

\subsection{Model of the UV Emission Lines in 2650-3050{\AA}: Extracting the Pure Mg II Profile}

We performed the spectral decomposition in 2650-3050{\AA} range in order to estimate the pure Mg II profile, which overlaps with the UV Fe II lines (Figure 1). The Mg II line was fitted with two Gaussian functions, the one that represents the line core and the other that fits the line wings \citep[see][]{Kovecevic-Dojcinovic_2015}, and a sum of those two components represents the broad Mg II line. The narrow Mg II line was not detected in analyzed spectra. The numerous Fe II lines in the range 2650-3050{\AA} were all fitted with the Fe II template presented in \citet{Popovic_2003} and \citet{Kovecevic-Dojcinovic_2015}. The Fe II lines in the 2650-3050{\AA} range were divided into four multiplets, and relative intensities of the lines within each multiplet were taken from literature. Finally, the Fe II template was described with six parameters: four parameters for the intensity, line width and shift. The example of the spectral decomposition in the 2650-3050{\AA} range with UV pseudo-continuum fit is shown in the UV part in Figure 1.

\subsection{Measuring the Spectral Parameters}
From the pure broad profiles of the H$\beta$ and Mg II lines, we measured the FWHM of these lines, as well as Full Width at 10\% of the Maximum (FW10\%M).
The asymmetries of these lines (intrinsic shifts) were measured at different levels of the line maximal intensity (at 50\%, z$_{50}$ and at 10\%, z$_{10}$), as the centroid shift with respect to the broad line peak \citep[see][their Fig 2]{Jonic_2016}. 
The luminosity of the continuum was calculated using the formula given in \citet{Peebles_1993} with adopted cosmological parameters of $\Omega_{M}$ = 0.3, $\Omega_{\Lambda}$ = 0.7, $\Omega_{k}$ = 0, and Hubble constant H$_o$ = 70 km s$^{-1}$ Mpc$^{-1}$. The virial M$_{\rm BH}$ for the UV parameters (FWHM Mg II, L$_{\lambda}(3000\mathring A)$) was calculated using the formula given in \citet{Wang_2009} (their E. (9), for $\gamma$ = 2).

\section{Results}

\subsection{Testing the Virialization of the Broad H$\beta$ and Mg II Lines}

If the emission gas in the BLR is virialized, one can expect to observe correlations between the widths and the gravitational redshifts of the BELs, which comes from the equations for the M$_{\rm BH}$ estimation by the virial method using the line width \citep[see][]{Zheng_Sulentic_1990,Peterson_2004}. The expected relation is: z$_G$ $\sim$ FWHM$^2$, i.e. log(z$_G$) $\sim$ log(FWHM) \citep[see][]{Jonic_2016}.  This correlation is expected since at the same distance from a BH the line is broadened by the gravitational rotational effect which depends from the distance as $\sim$ $\sqrt{\frac{M}{r}}$, and by the gravitational shift (red-shifting of photons) that also depends from the distance as $\sim$ $\frac{M}{r}$ \citep[see][]{Popovic_1995}.

We investigated correlation between the intrinsic shifts, as indicators of the gravitational redshifts, and the widths of the H$\beta$ or the Mg II lines, at different levels (50\% and 10\%) of their maximal intensity, for the sample of 123 Type 1 AGNs with the red asymmetry in BELs. We have found that the width of the H$\beta$ line is well correlated with the line's intrinsic shift measured at the 50\% and at the 10\% of the maximal intensity. However, in the case of the Mg II line, the correlation between the Mg II width and intrinsic shift is detected only at the 50\% of the line maximal intensity, whereas at the 10\% of the line maximal intensity an anti-correlation is seen \citep[for more detail see][]{Jonic_2016}. 
\vspace{-2ex}
\begin{figure}[H]
\begin{center}
\centering\includegraphics[width=17.5cm]{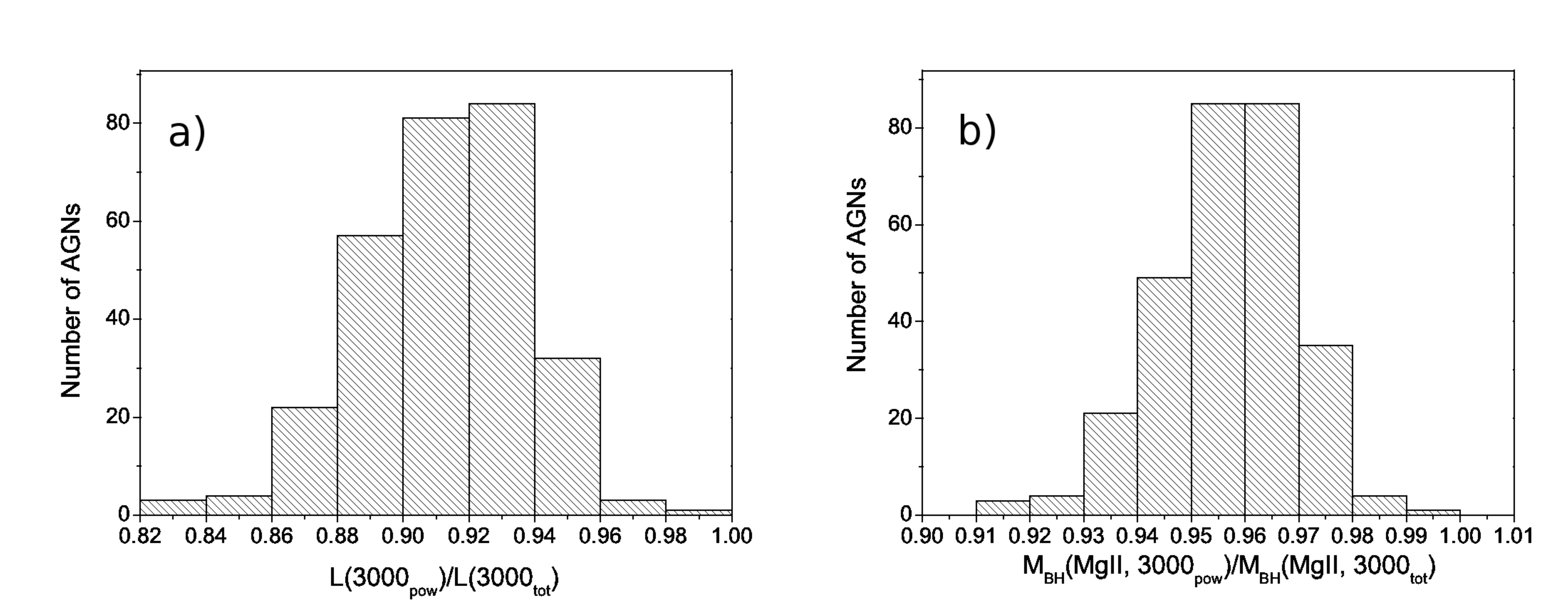}
\caption{a. The ratio of the continuum L$_{\lambda}(3000\mathring A)$ before (L$(3000\mathring A)_{\rm tot}$)) and after (L$(3000\mathring A)_{\rm pow}$) the Balmer continuum subtraction; b. The ratio of the M$_{\rm BH}$ estimate before (M$_{\rm BH}$ (L$_{\rm tot}$)) and after (M$_{\rm BH}$ (L$_{\rm pow}$)) the Balmer continuum subtraction.}
\label{fig:2}
\end{center}
\end{figure}

Note here that the literature in the field presents comparisons on the use of the optical and the UV lines, like H$\beta$ and Mg II, as viral estimators of the M$_{\rm BH}$ in AGN, pointing out that in some cases Mg II is more reliable as M$_{\rm BH}$ estimator then H$\beta$ \citep[e.g.][and references therein]{Marziani_2013}. However, it has been shown that the Mg II wings are emitted from the weakly gravitationally bounded gas \citep[see][]{Kovecevic-Dojcinovic_2015,Jonic_2016}, i.e. the assumption of virialization may be problem in using Mg II line as the BH mass estimator.
\nohyphens{
\subsection{Influence of the Blamer Continuum to the M$_{\rm BH}$ Estimation Using the UV Spectral Parameters}
}
We calculated the luminosity at 3000{\AA} before ( L$_{\lambda}(3000\mathring A)_{\rm tot}$) and after the Balmer continuum (L$_{\lambda}(3000\mathring A)_{\rm pow}$) contribution was subtracted. In the sample of 287 Type 1 AGN spectra, the Balmer continuum contributes to the continuum L$_{\lambda}(3000\mathring A)$ on average for $\sim$ 10\%, with the maximal value of 18\%. The ratios of the L$_{\lambda}(3000\mathring A)$, with and without the Balmer continuum contribution, are shown on the histogram (Figure 2a). 
\vspace{-7ex}
\begin{figure}[H]
\begin{center}
\centering\includegraphics[width=18cm]{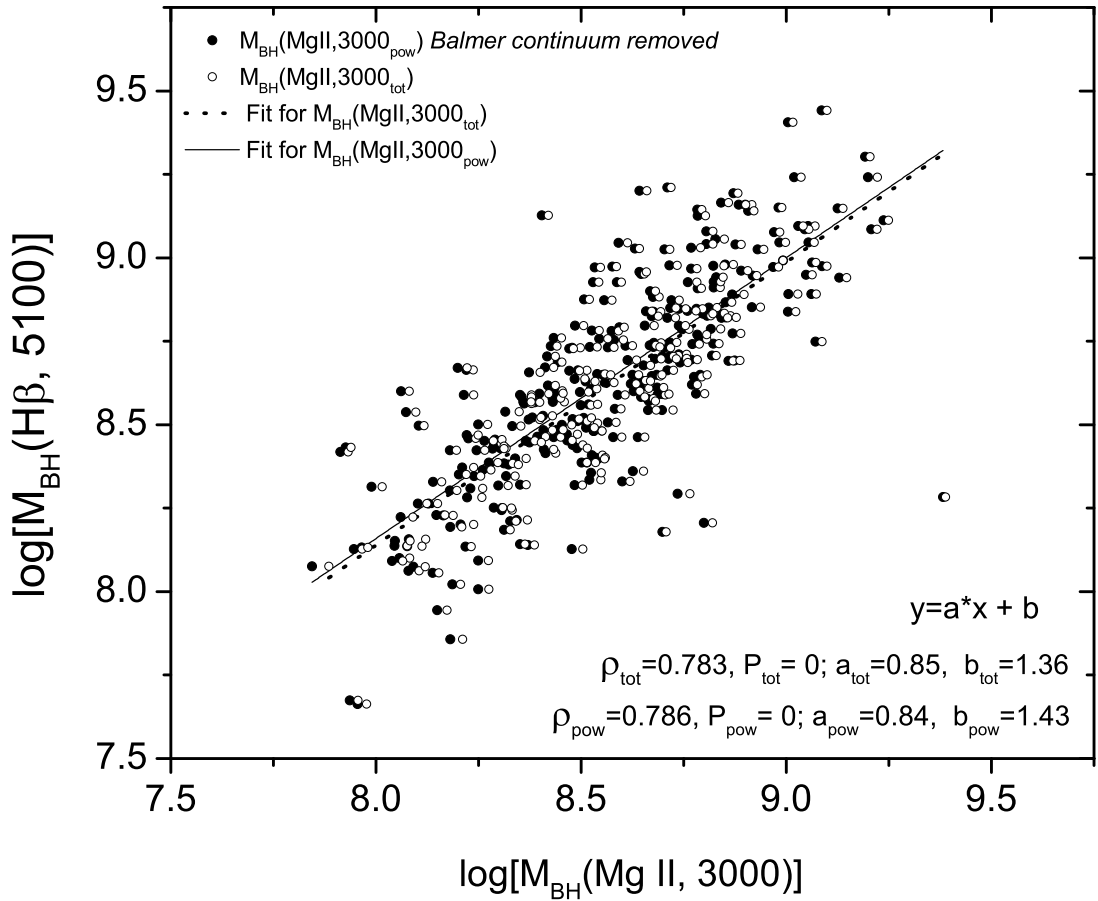}
\caption{The relation of the M$_{\rm BH}({\rm H}\beta$,L$_{\lambda}(5100\mathring A))$ vs. M$_{\rm BH}({\rm MgII}$,L$_{\lambda}(3000\mathring A))$ before (index 'tot', open circles) and after the Balmer continuum subtraction (index 'pow', full circles). The corresponding correlation coefficients, P-values, and coefficients of the liner best-fit are also given.}
\label{fig:3}
\end{center}
\end{figure}
\vspace{-3ex}

The majority of the equations for the M$_{\rm BH}$ estimation using the UV parameters neglect the contribution of the Balmer continuum \citep[see e.g.][]{McJure_Jarvis_2002,Vestergaard_Osmer_2009}. \citet{Wang_2009} give the relation for the assessment of M$_{\rm BH}$ using the FWHM Mg II and L$_{\lambda}(3000\mathring A)$, assuming that after subtraction of the Balmer continuum, the L$_{\lambda}(3000\mathring A)$ is the pure power law continuum, therefore we then compared M$_{\rm BH}$ calculated with and without consideration of the Balmer continuum contribution. 

We got that the Balmer continuum increased M$_{\rm BH}$ on average for $\sim$ 5\% (0.02 dex), with the maximal value of the M$_{\rm BH}$ overestimation up to 10\% (0.04 dex). The ratio of the M$_{\rm BH}$ estimates before and after the Balmer continuum subtraction is shown in Figure 2b. The comparison of M$_{\rm BH}$ values before (index 'tot') and after the Balmer continuum subtractions (index 'pow') are also given in Figure 3, where the correlation coefficients $\rho$ ($\rho_{tot}=0.783$ and $\rho_{pow}=0.786$), P-values (P$_{tot}$=0 and P$_{pow}$=0) and linear best-fit (y=a*x+b) coefficients are given: slope a, (a$_{tot}$=0.85 and a$_{pow}$=0.84) and y-intercept b, (b$_{tot}$=1.36 and b$_{pow}$=1.43). It can be seen that the influence of the Balmer continuum to the M$_{\rm BH}$ estimation is so small that it barely changes the correlation coefficient ($\rho$) or the coefficients of the linear best-fit (a, b) between the M$_{\rm BH}$ estimated with the optical and the one estimated with the UV parameters. Also, it seams that removing the Balmer continuum does not affect the outliers in this relationship.
\vspace{2ex}
\section{Conclusions}

Here we have used the sample of the SDSS Type 1 AGN spectra to compare the most frequently used emission lines for the M$_{\rm BH}$ estimation, H$\beta$ (in the optical band) and Mg II (in the UV band), in order to assess which line is better virial estimator and thus more convenient for that purpose. We investigated how the Balmer continuum affect the BH mass estimation using the UV parameters. 

From our investigation we can outline the following conclusions:
\begin{enumerate}
\item The H$\beta$ line is a more reliable virial estimator than the Mg II line, since the expected linear relationship between logarithms of the widths (influenced by the Keplerian motion) and red asymmetries (caused by the gravitational redshift) was evidenced for both lines when measured at the 50\% of the line maximal intensities, but when measured in the line wings (at the 10\% of the line maximal intensities) the expected relationship was present only for H$\beta$ \citep[see][]{Jonic_2016}. 

\item The disregard of the Balmer continuum emission, in the case of the M$_{\rm BH}$ estimation using the UV parameters (Mg II, L$_{\lambda}(3000\mathring A)$), causes the overestimation of the M$_{\rm BH}$ on average for $\sim$  5\% (0.02 dex) and up to 10\% (0.04 dex).
\end{enumerate}

At the end, let us note that similar investigation should be performed on the sample where more reliable methods for mass measurements (as e.g. reverberation) should be applied to explore the influence of the Balmer continuum and this we postpone for our future work. Moreover, some additional effects (as e.g. relativistic jets) can significantly affect line profiles, i.e. the radio loudness which can indicate the presence of relativistic jets, therefore in the future work radio properties of the sample should also be explored.

\section*{Acknowledgments}
The work is a part of the project 176001 financed by the Ministry of Education, Science, Technology and Development, Republic of Serbia.

\bibliographystyle{rusnat}

\bibliography{mybib3}


\end{document}